\documentclass[5p]{elsarticle}

\usepackage{graphicx}
\usepackage{amssymb,amsthm}
\usepackage{color}

\begin{document}

\begin{frontmatter}

\title{Chimera-like in a neuronal network model of the cat brain}

\author{M. S. Santos$^1$, J. D. Szezech Jr$^{1,2}$, F. S. Borges$^3$,
K. C. Iarosz$^{3,4}$, I. L. Caldas$^3$, A. M. Batista$^{1,2,3,4}$,
R. L. Viana$5$, J. Kurths$^{4,6}$}
\address{$^1$P\'os-Gradua\c c\~ao em Ci\^encias, Universidade Estadual de Ponta
Grossa, Ponta Grossa, PR, Brazil.}
\address{$^2$Departamento de Matem\'atica e Estat\'istica, Universidade
Estadual de Ponta Grossa, Ponta Grossa, PR, Brazil.}
\address{$^3$Instituto de F\'isica, Universidade de S\~ao Paulo, S\~ao Paulo,
SP, Brazil.}
\address{$^4$Institute for Complex Systems and Mathematical Biology, Aberdeen,
Scotland.}
\address{$^5$Departamento de F\'isica, Universidade Federal do Paran\'a,
Curitiba, PR, Brazil.}
\address{$^6$Department of Physics, Humboldt University, Berlin, Germany; and
Potsdam Institute for Climate Impact Research, Potsdam, Germany.}

\cortext[cor]{Corresponding author: antoniomarcosbatista@gmail.com}

\date{\today}

\begin{abstract}
Neuronal systems have been modelled by complex networks in different description
levels. Recently, it has been verified that networks can simultaneously exhibit
one coherent and other incoherent domain, known as chimera states. In this
work, we study the existence of chimera-like states in a network considering the
connectivity matrix based on the cat cerebral cortex. The cerebral cortex of
the cat can be separated in 65 cortical areas organised into the four cognitive
regions: visual, auditory, somatosensory-motor and frontolimbic. We consider a
network where the local dynamics is given by the Hindmarsh-Rose model. The
Hindmarsh-Rose equations are a well known model of neuronal activity that has
been considered to simulate membrane potential in neuron. Here, we analyse
under which conditions chimera-like states are present, as well as the affects
induced by intensity of coupling on them. We identify two different kinds of
chimera-like states: spiking chimera-like with desynchronised spikes, and
bursting chimera-like with desynchronised bursts. Moreover, we find that
chimera-like states with desynchronised bursts are more robust to neuronal
noise than with desynchronised spikes.
\end{abstract}

\begin{keyword}
chimera-like \sep neuronal network \sep noise
\PACS 05.45.Pq \sep 87.19.lj 
\end{keyword}

\end{frontmatter}


\section{Introduction}

The mammalian brain has neuronal mechanisms that give support to various
anatomically and functionally distinct structures \cite{barton00}. Mammals have
the most complex brains of all vertebrates, which vary in size by a factor
of $10^5$ \cite{willemet12}. Such a brain is arranged according to not only
interacting elements on different levels, but also of different
interconnections and functions \cite{zemanova06}. For instance, the cat has
approximately $10^9$ neurons in the brain and $10^{13}$ synapses 
\cite{ananthanarayanan2009,binzegger2004}, while the human brain has
approximately $10^{11}$ neurons and $10^{14}$ synapses \cite{williams1988}. 

One of the mammalian brain connectivity studies that has received considerable
attention is the connectivity in the cat cerebral cortex 
\cite{scannell95,beul15}. Scannell and Young \cite{scannell93} reported the
connectional organisation of neuronal systems in the cat cerebral cortex.
They arranged the cortex in four cognitive regions or connectional groups of
areas: visual, auditory, somatosensory-motor, and frontolimbic. Regarding this
realistic neuronal network, it was studied the relationship between structural
and functional connectivity at different levels of synchronisation
\cite{zhou07}. Lameu et al. analysed bursting synchronisation \cite{lameu16a}
and suppression of phase synchronisation \cite{lameu16b} in network based on
cat's brain. They verified that the delayed feedback control can be an
efficient method to have suppress synchronisation.

We focus on the existence of chimera-like states in a neuronal network model 
based on the cat cerebral cortex. The chimera states are spatiotemporal 
patterns in which coherent and incoherent domains mutually coexist 
\cite{kuramoto02,abrams04}. There are many studies about these patterns
\cite{abrams08,ujjwal16}. Omel'chenko et al. observed chaotic motion of the
chimera's position along arrays of nonlocally coupled phase oscillators,
where the chimera states have no artificially imposed symmetry
\cite{omelchenko10}. Chimera and phase-cluster states in populations of
coupled chemical oscillators were studied by Tinsley et al. \cite{tinsley12}.
Considering coupled Belousov-Zhabotinsky oscillators, they verified that
chimera lifetime grows approximately exponentially with system size. There
are also experimental investigations about chimera states, e.g., in
\cite{gambuzza14} demonstrated the existence of these patterns in an open chain
of electronic circuits with neuron-like spiking dynamics. In addition, Martens
et al. \cite{martens13} showed the appearance of chimera states in
experiments with mechanical oscillators coupled in a hierarchical network.

Coexistence of states was reported in several animals that exhibit conflict
between sleep and wakefulness, where one cerebral hemisphere sleeps and the
other stays in an awake condition \cite{rattenborg00}. Recently, Andrzejak et
al. \cite{andrzejak16} demonstrated analogies between chimera state
collapses and epileptic seizures. In neuronal systems, chimera states were
found in a network of coupled Hodgkin-Huxley equations \cite{sakaguchi06}, and
also in the C. Elegans brain network by coupling Hindmarsh-Rose equations
\cite{hizanidis15,hizanidis16}. 

We build a neuronal network according to the matrix of corticocortical 
connections in the cat \cite{scannell95,scannell93}. This topological
connectivity is more complex than the matrix description from C. Elegans due to
the fact that it is composed of cortical areas and axonal projections between
them. In each cortical area we consider the neuron model proposed by Hindmarsh
and Rose (HR) \cite{hindmarsh84}, and the axonal densities are considered as
the connections between the areas. This neuronal network is able to reproduce
EEG-like oscillations \cite{schmidt10}. The HR neuron model can reproduce
neuronal activities such as regular or chaotic spikes, and regular or chaotic
bursting \cite{storace08}. Baptista et al. \cite{baptista10} analysed a HR
neuronal network on the rate of information and synchronous behaviour.
Hizanidis et al. \cite{hizanidis14} observed chimera-like states in nonlocally
coupled HR neuron models, where each neuron is connected with its nearest
neighbours on both sides. In this work, we consider unidirectional connections
both inside each connectional group of area (intra) and between groups of areas
(inter). Unidirectional connections can be related to chemical synapses
\cite{eccles82}. As a result, we verify the existence of two kinds of
chimera-like: spiking chimera-like (SC) with desynchronised spikes, and
bursting chimera-like (BC) with desynchronised bursts. Moreover, we include a
neuronal noise by adding a random term in the external current. This way, we
demonstrate that BC is more robust to noise than SC.

Firstly, we introduce the neuronal network described by a coupled HR neuronal
model. Then, we discuss our results about the existence of two kinds of
chimera-like states and noise robustness. Finally, we draw our conclusions.


\section{Neuronal network}

\begin{figure}[hbt]
\centering
\includegraphics[width=0.95\linewidth]{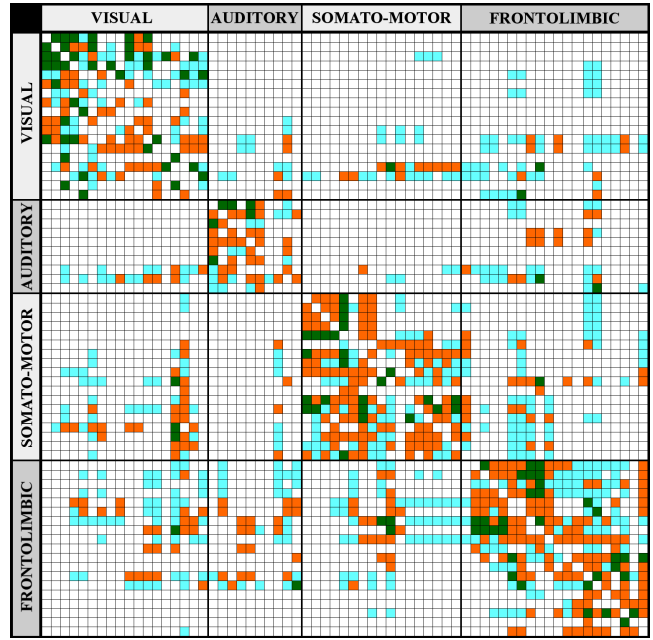}
\caption{(Colour online) Density of connections between cortical areas 
classified as absent of connection (white), sparse or weak (cyan), intermediate 
(orange), and dense or strong (green).}
\label{fig1}
\end{figure}

We consider a neuronal network composed for coupled HR according to the matrix
that describes the corticocortical connectivity of the cat brain, obtained by
Scannel et. al \cite{scannell95}. Fig. \ref{fig1} shows the matrix of
corticocortical connections in the cat brain with 1139 connections between 65
cortical areas which is organised into four cognitive regions: visual, auditory,
somatosensory-motor and frontolimbic. In Ref. \cite{scannell95}, the matrix
was constructed by means of connections weighted $0$, $1$, $2$, or $3$, where
$0$ corresponds to absent of connections (white), $1$ are sparse or weak
(cyan), $2$ are intermediate (orange), and connections weighted $3$ are dense
or strong (green).

The dynamic behaviour of the HR network if governed by the following equations
\begin{eqnarray}
\dot{x_{j}} & = & y_{j}-x_{j}^{3}+bx_{j}^{2}+I_{j}-z_{j}-
\frac{\alpha}{n^{'}}\sum_{k=1}^{N} G^{'}_{j,k}\Theta(x_k)-\nonumber\\
& & \frac{\beta}{n^{''}}\sum_{k=1}^{N}G^{''}_{j,k}\Theta(x_k),\nonumber\\
\dot{y_{j}} & = & 1-5x_{j}^{2}-y_{j},\nonumber\\
\dot{z_{j}} & = & \mu\left[s\left(x_{j}-x_{\rm rest}\right)-z_{j}\right],
\label{eq01}
\end{eqnarray}
where $\Theta(x_k)=(x_j-x_{\rm rev})[1+e^{-\lambda(x_{k}-\theta)}]^{-1}$, $x_j$ is the 
membrane potential, $y_j$ is related to the fast current (Na$^+$ or K$^+$), 
$z_j$ is associated with the slow current (Ca$^{2+}$), $b$ controls the spiking 
frequency, $I_{j}$ corresponds to membrane input current ($1\leq\,j\,\leq N$), 
$x_{\rm rev}$ is the reversal potential, $\lambda$ and $\theta$ are sigmoidal 
function parameters, $\mu$ is responsible for the speed of variation of $z$, 
$s$ governs adaptation, $x_{\rm rest}$ is the resting potential, $\alpha$ is the 
intra connection strength, $G^{'}_{j,k}$ is the connection matrix of connections
inside cortical areas (intra), $\beta$ is the inter connection strength, 
and $G^{''}_{j,k}$ is the connection matrix of connections between cortical areas
(inter). We fix $b=3.2$, $I_{j}=I_{0}=4.4$, $N=65$ cortical areas,
$x_{\rm rev}=2$, $\lambda=10$, $\theta=-0.25$, $\mu=0.01$, $s=4.0$, and 
$x_{\rm rest}=1.6$ \cite{hizanidis16,storace08}. The elements of the matrices
$G^{'}_{j,k}$ and $G^{''}_{j,k}$ are $0$ (absent of connection), $1/3$ (weak),
$2/3$ (intermediate), or $1$ (strong) according to the cat matrix (Fig.
\ref{fig1}).

\begin{figure}[hbt]
\centering
\includegraphics[width=0.95\linewidth]{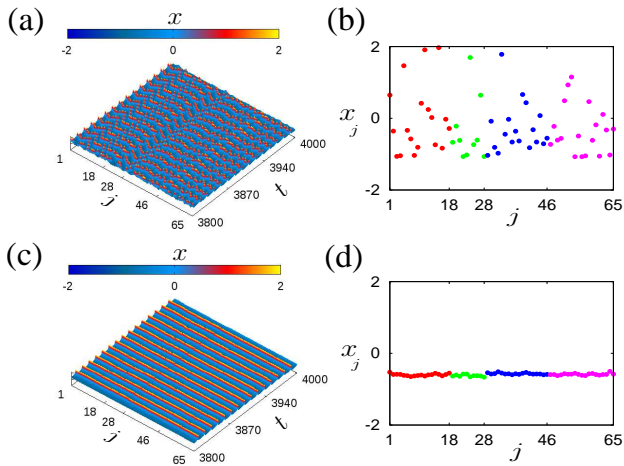}
\caption{(Colour online) Space-time plots (left) and snapshot of the variable
$x$ (right). (a) and (b) exhibit desynchronous behaviour for $\alpha=0.001$ and 
$\beta=0.001$. (c) and (d) show synchronous behaviour for $\alpha=0.21$ and 
$\beta=0.04$.}
\label{fig2}
\end{figure}

Fig. \ref{fig2} shows space-time plots (left) and snapshot of the variable $x$
(right). We use, in the snapshot, a colour for each cortical areas: visual in
red, auditory in green, somatosensory-motor in blue, and frontolimbic in 
magen\-ta. In Fig. \ref{fig2}a for $\alpha=0.001$ and $\beta=0.001$, the network
has a desynchronous behaviour, namely it is not possible to observe a
synchronised firing pattern. Consequently, the dynamics is spatially
incoherent, as shown in Fig. \ref{fig2}b. For $\alpha=0.21$ and $\beta=0.04$,
there is a clear synchronised firing pattern (Fig. \ref{fig2}c) and the network
displays a spatially coherent dynamics (Fig. \ref{fig2}d).

\section{Chimera-Like states}

\begin{figure}[hbt]
\centering
\includegraphics[width=0.95\linewidth]{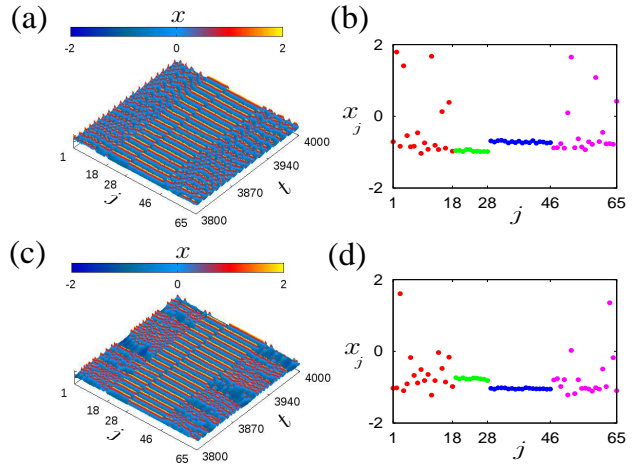}
\caption{(Colour online) Space-time plots (left) and snapshot of the variable
$x$ (right). (a) and (b) exhibit SC for $\alpha=0.7$ and $\beta=0.08$. (c)
and (d) show BC for $\alpha=1.5$ and $\beta=0.1$.}
\label{fig3}
\end{figure}

One fact that has been verified is the coexistence of coherence and incoherence
structures in networks \cite{santos15}. This phenomenon in spatiotemporal 
dynamical systems is so-called chimera states \cite{kuramoto02,abrams04}. Fig. 
\ref{fig3} shows space-time plots (left) and snapshots (right) for different 
values of $\alpha$ and $\beta$ in that it is possible to identify chimera-like
states. In Figs. \ref{fig3}a and \ref{fig3}b for $\alpha=0.7$ and $\beta=0.08$,
the auditory (green) and the somatosensory-motor area (blue) have synchronous 
behaviours, namely they exhibit spatially coherent dynamics, while the visual 
(red) and frontolimbic (magenta) areas are spatially incoherent with spikes
patterns. This chimera-like is SC because the incoherent intervals are
characterized by desynchronised spikes. For $\alpha=1.5$ and $\beta=0.1$ (Figs.
\ref{fig3}c and \ref{fig3}d), the auditory (green) and somatosensory-motor
(blue) areas display synchronised patterns, whereas the other areas show
incoherent structures. In this case, we verify BC, that is when desynchronised
bursts are observed in the incoherent intervals. The transition between
bursting and spiking resembles a continuous interior crisis 
\cite{miranda03,innocenti07}. The interior crisis in the HR is a type of 
chaos-chaos transition, where it is observed changes of the size of the chaotic
attractor due to collision between an unstable periodic orbit and a chaotic 
attractor.

In this work, we propose to use the recurrence plot as diagnostic tool to
identify chimera-like in HR neurons coupled according to the cat matrix. To
obtain this diagnostic, it is necessary to calculate the neuronal phase that is
defined as
\begin{equation}
\phi_{j}\left(t\right)=2\pi k+2\pi\frac{t-t_{k}}{t_{k+1}-t_{k}},
\end{equation}
where $t_{k}$ is the firing time of the $j$-th neuron. Then, the recurrence
plot is given by \cite{santos15}
\begin{equation}
RP_{i,j}=\Phi(\epsilon-|\phi_i-\phi_j|),
\end{equation}
where $\epsilon=0.3$. Chimera-like states are found when there are blocks
in the main diagonal and sparse points in the $RP_{ij}$. We consider that
chimera states occur when the sizes of the blocks are larger than $50\%$ of the
total number of areas in each region. When there are only blocks and no sparse
points, the network exhibits synchronized states (SI). Whereas, the incoherent
states (IN) are identified when the blocks sizes are less than $50\%$. 

As a criterium to differentiate SC and BC we use the spiking time variance, 
that is given by
\begin{equation}
\sigma=\langle\Delta t_k^2\rangle-\langle\Delta t_k\rangle^2,
\end{equation}
where $\Delta t_k=t_k^{(\tau+1)}-t_k^{(\tau)}$, if the first spike occurs at time
$t_k^{(\tau)}$ then the next one at time $t_k^{(\tau+1)}$, and the symbol 
$\langle . \rangle$ refers to the mean. For $\sigma\leq 10$ the network exhibits
SC and for $\sigma>10$ it presents BC.

\begin{figure}[hbt]
\centering
\includegraphics[width=0.95\linewidth]{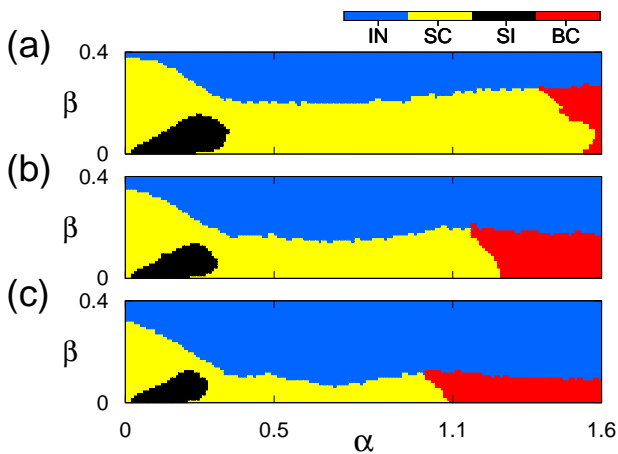}
\caption{(Colour online) Parameter spaces $\beta\times\alpha$ for $b=3.2$,
(a) $I_0=4.6$, (b) $I_0=4.4$, and (c) $I_0=4.2$. We see the regions for
{\color{yellow}SC}, {\color{red} BC}, SI, and {\color{blue}IN}.}
\label{fig4}
\end{figure}

Figs. \ref{fig4}a, \ref{fig4}b and \ref{fig4}c show in the parameter space
$\beta\times\alpha$ the regions for SC, BC, SI, and IN. In the parameter space
$\beta\times\alpha$, each point is computed by means of the average of $100$
initial conditions, where the initial conditions are randomly distributed in
the intervals $x_j\in [-2,2]$, $y_j\in[0,0.2]$ and $z_j \in[0,0.2]$. In these
intervals, the individual Hindmarsh-Rose neuron model exhibits spiking
behaviour. This way we can identify the regions where chimera-like states
appear by examining the recurrence plot. The yellow and red regions correspond
to SC and BC, respectively, that are characterised by means of the spiking time
variance. In the parameter space, the SC region is larger than the BC region.
The black region exhibits SI and in the blue region we see IN. In our
simulations, we observe that the IN region increases when $I_0$ decreases from
$4.6$ to $4.2$ and, as a consequence, the SC and BC regions have a significant
decrease in their sizes. In addition, for small $\beta$ values we verify that
SC changes to BC.

\section{Matching index}

Some papers showed that the cat array has a high hierarchical level 
\cite{zhou07,sporns07}. This enables the formation of hierarchical 
synchronization \cite{zamora10,gomez-gardenes10,stroud15}. We verify that the
coherent regions in the chimera-like states are related to synchronisation
between the neurons of the same connectional group of areas, mainly in auditory
and somatosensory-motor.

Aiming to understand the formation of chimera-like states, we calculate the 
matching index (MI) of each two nodes $i$ and $j$ of the matrix showed in Fig. 
\ref{fig1}. The matching index of two nodes $i$ and $j$ is the overlap of their
neighborhoods and is given by \cite{hilgetag02}
\begin{eqnarray}
{\rm MI}_{ij}= A_{ij} + \sum_{n=1}^{N} A_{in} A_{jn},
\label{eqMI}
\end{eqnarray}
where $A_{ij}$ is the adjacency matrix with elements equal to $1$ or $0$ 
according to whether $i$ and $j$ are connected or not. The MI is normalised 
dividing each element of the matrix by $k_i+k_j-{\rm MI}_{ij}$, where $k_i$ is 
the degree of the node $i$. ${\rm MI}_{ij}$ is equal to $0$ or $1$ if all inputs
to $i$ and $j$ come from entirely different areas or $i$ and $j$ receive input
completly from the same cortical areas, respectively.

\begin{figure}[hbt]
\centering
\includegraphics[width=0.95\linewidth]{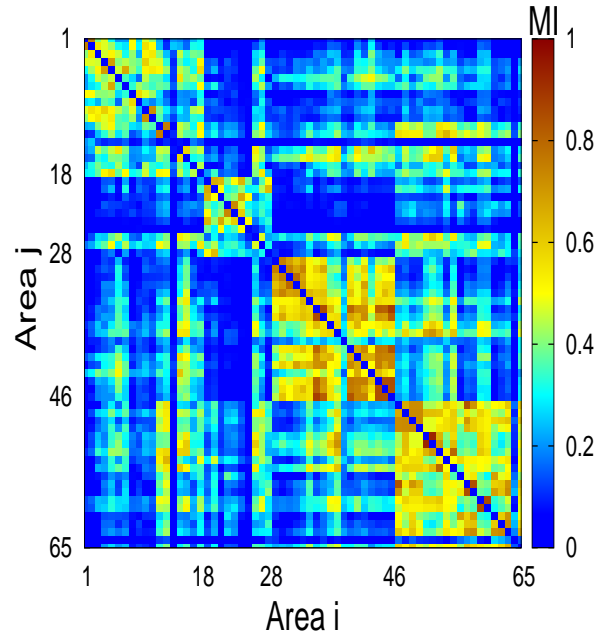}
\caption{(Colour online) Normalised matching index $MI_{ij}$ for all areas from
the anatomical connectivity matrix, where self-matching $MI_{ii}$ is ignored.}
\label{fig5}
\end{figure}

Fig. \ref{fig5} exhibits the normalised MI matrix from the anatomical 
connectivity. We see high MI values for the areas within the connectional group
of visual, auditory, somatosensory-motor, and frontolimbic (internal MI).
Areas in different group of areas (external) have heterogeneous MI values
\cite{zhou07}.

The average of the normalised MI is $0.26$ for the whole matrix. The internal 
MI values are $0.35$, $0.385$, $0.54$, and $0.454$ for the connectional visual,
auditory, somatosenso\-ry-motor, and frontolimbic group of areas, respectively.
Furthermore, the auditory region has a higher value of MI ($= 0.512$) when 
it is only considered the areas between $19-24$. In Refs. 
\cite{zhou07,gomez-gardenes10} was showed that high values of MI can cause 
cluster synchronisation. 

In our simulations, we verify that cluster synchronisation helps the appearance
of coherent regions in the network, and as a consequence if incoherent regions
mutually coexist, the network exhibits chimera-like states.


\section{Neuronal noise}

A relevant brain feature is noise, that can affect the transmission of signals 
and neuronal function \cite{faisal08}. There are many sources of noise in the 
brain, such as from genetic processes, thermal noise, ionic channel 
fluctuations, and synaptic events \cite{destexhe12}. Serletis et al. 
\cite{serletis13} studied phase synchronisation of neuronal noise in mouse 
hippocampal. They reported from multi-spatial recordings that noise activity
has a great influence on neurodynamic transitions in the healthy and epileptic
brain.

We study the influence of neuronal noise on the chime\-ra-like states in the cat
brain. To do that, we consider an input current in Eq. 1 that is given by
\begin{equation}
I_{j}=I_{0}+\delta \Psi_{j},
\label{eq001}
\end{equation}
where $\delta$ is the amplitude and $\Psi_{j}$ is a normal distribution with
mean $0$ and variance $1$ (Gaussian noise). For $\delta=0$ the isolated
neuron exhibits spiking pattern. There are many studies that consider neurons
under the influence of Gaussian noise \cite{lindner03}. Loos et al.
\cite{loos16} investigated chimera states under the influence of Gaussian noise
in ring networks of Stuart-Landau oscillators. They showed that chimera death
patterns persist under the impact of noise, and they also verified that the
range of coupling parameters where chimera states occur is reduced with
increasing noise amplitude.

\begin{figure}[hbt]
\centering
\includegraphics[width=0.95\linewidth]{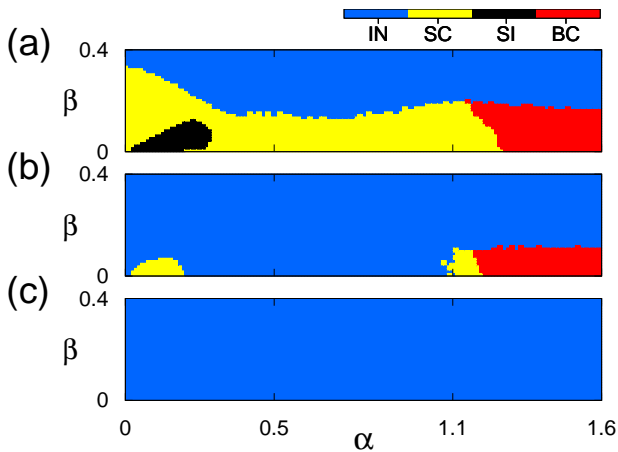}
\caption{(Colour online) Parameter spaces $\beta\times\alpha$ for (a) 
$\delta=0.04$, (b) $\delta=0.16$, and (c) $\delta=0.25$.}
\label{fig6}
\end{figure}

Fig. \ref{fig6} shows the parameter spaces $\beta\times\alpha$ under an 
external current with noise (Eq. \ref{eq001}). For a small noise amplitude we 
do not observe significant alteration in the regions, as shown in Fig.
\ref{fig6}a for $\delta=0.04$. However, when $\delta$ is increased to $0.16$ it
is possible to verify the destruction not only of synchronous behaviour, but 
also of chimera-like states (Fig. \ref{fig6}b). In Figs. \ref{fig6}c, where we 
consider $\delta=0.25$, synchronisation and chimera-like states are completely 
suppressed.

\begin{figure}[hbt]
\centering
\includegraphics[width=0.95\linewidth]{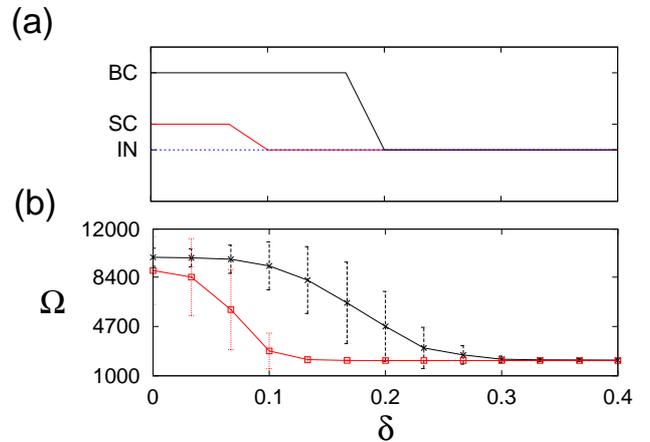}
\caption{(Colour online) Noise robustness as a function of $\delta$. (a) for
desynchronised bursts represented by the black line ($\alpha=1.5$ and
$\beta=0.1$) and desynchronised spikes represented by the red line
($\alpha=0.7$ and $\beta=0.08$). (b) average chimera-like lifetime and the
standard deviation calculated by means of 400 initial conditions.}
\label{fig7}
\end{figure}

We examine the robustness to noise of chimera-like states with spiking and
bursting structures. To do that, we vary the amplitude $\delta$ and calculate 
the recurrence plot and the spiking variance. The results in Fig. \ref{fig7} 
correspond to $\alpha=0.7$ and $\beta=0.08$ into SC region (red line), and 
$\alpha=1.5$ and $\beta=0.1$ into BC region (black line). In Fig. \ref{fig7}a,
we see that for $\delta \gtrsim 0.1$ and $\delta \gtrsim 0.2$ the noise 
suppresses the SC and BC, respectively. Therefore, chimera-like states with 
desynchronised bursts are more robust to noise than with desynchronised spikes.
We also calculate the average chimera-like lifetime $\Omega$, as shown in Fig. 
\ref{fig7}b, considering $400$ different initial conditions. The noise decreases
the average chimera-like lifetime of both SC and BC.

\section{Conclusions}
 
We study a neuronal network composed by coupled HR model according to the
connectivity matrix of the cat cerebral cortex. We consider unidirectional
connections between neurons in the same area (intra) and between neurons in
different areas (inter). This kind of connections can be associated with
chemical synapses. The isolated HR neuron model has a regular spiking, but HR
neurons in the network can exhibit bursting behaviour due to the coupling.

Varying the strength couplings, the network can exhibit synchronous and 
desynchronous behaviour. We also observe spatiotemporal patterns in that 
coherent and incoherent structures coexist, so-called chimera-like states. As 
diagnostic tool we propose to use the recurrence plot to identify chimera-like
states in the neuronal network model of the cat brain. As a result, we verify
that the network displays chimera-like states where the incoherent structures
can be composed by desynchronised spikes or desynchronised bursts. SC occurs
for small intra coupling strength, while BC appears for large intra coupling
strength, and both for small inter coupling strength. 

There are many sources of noise in the brain. With this in mind, we consider
a noise in the external current to analyse effects on the chimera-like states.
For small noise there is not significant changes in the parameter space
($\alpha\times\beta$) related to chimera-like, however, when the noise amplitude
increases, chimera-like suppression is observed with the neuronal network
exhibiting a desynchronous behaviour. In addition, we verify that BC is more 
robust to noise than SC, and the noise decreases the average chimera-like
lifetime. 

In future works, we plan a deeper analysis of the root cause of the 
chimera-like phenomenon in the cat network. It will be analysed the instability
of synchronisation in terms of the master stability function
\cite{stroud15,barahona02}.


\section*{Acknowledgments}
This study was possible by partial financial support from the following 
Brazilian government agencies: CNPq (154705/2016-0), CAPES, Funda\c c\~ao
Arauc\'aria, and FA\-PESP (2016/16148-5, 2015/07311-7, and 2011/19296-1), and
IRTG 1740/TRP 2011/50151-0 funded by the DFG/ FAPESP.  


\begin{frame}

\end{frame}
\end{document}